\documentclass[12pt]{article}
\usepackage{graphics}
\usepackage{amssymb}
\usepackage{amsmath}

\newcommand{\rf}[1]{(\ref{#1})}
\newcommand{\beq}{\begin{equation}}
\newcommand{\eeq}{\end{equation}}
\newcommand{\bea}{\begin{eqnarray}}
\newcommand{\eea}{\end{eqnarray}}

\newcommand{\e}{\mbox{e}}
\renewcommand{\d}{\mbox{d}}

\newcommand{\lam}{\lambda}

\newcommand{\n}{\nu}
\newcommand{\m}{\mu}

\newcommand{\Om}{\Omega}
\newcommand{\del}{\delta}

\newcommand{\oh}{\frac{1}{2}}

\newcommand{\dg}{\dagger}

\newcommand{\ra}{\rangle}
\newcommand{\rra}{\right\rangle}
\newcommand{\la}{\langle}
\newcommand{\lla}{\left\langle}
\newcommand{\prt}{\partial}

\newcommand{\cD}{{\cal D}}

\newcommand{\tW}{{\tilde{W}}}

\newcommand{\tG}{{\tilde{G}}}

\newcommand{\tH}{{\tilde{H}}}
\newcommand{\tpsi}{{\tilde{\psi}}}

\newcommand{\hH}{{\hat{H}}}

\newcommand{\hp}{{\hat{p}}}
\newcommand{\hx}{{\hat{x}}}

\newcommand{\hG}{{\hat{G}}}
\newcommand{\hR}{{\hat{R}}}

\newcommand{\sla}{\sqrt{\lam}}
\newcommand{\sOm}{\sqrt{\Om}}

\begin{document}

\begin{center}

{ \Large \bf Stochastic quantization and the role of time\\ in quantum gravity}

\vspace{30pt}

{\sl J.\ Ambj\o rn}$\,^{a,b}$, {\sl R.\ Loll}$\,^{b}$,
 {\sl W.\ Westra}$\,^{c}$ and
{\sl S.\ Zohren}$\,^{d,e}$

\vspace{24pt}

{\footnotesize

$^a$~The Niels Bohr Institute, Copenhagen University\\
Blegdamsvej 17, DK-2100 Copenhagen \O , Denmark.\\
{ email: ambjorn@nbi.dk}\\

\vspace{10pt}

$^b$~Institute for Theoretical Physics, Utrecht University, \\
Leuvenlaan 4, NL-3584 CE Utrecht, The Netherlands\\
{ email: r.loll@uu.nl}\\

\vspace{10pt}

$^c$~Department of Physics, University of Iceland,\\
Dunhaga 3, 107 Reykjavik, Iceland\\
{ email: wwestra@raunvis.hi.is}\\

\vspace{10pt}

$^d$~Mathematical Institute, Leiden University,\\
Niels Bohrweg 1, 2333 CA Leiden, The Netherlands\\
{email: zohren@math.leidenuniv.nl}\\ 

\vspace{10pt}

$^e$~Department of Statistics, Sao Paulo University,\\
Rua do Matao, 1010, 05508-090, Sao Paulo, Brazil

}
\vspace{48pt}

\end{center}

%\addtolength{\baselineskip}{0.20\baselineskip}
%\vspace{2cm}

\begin{center}
{\bf Abstract}
\end{center}

\noindent We show that the noncritical string field theory developed from 
two-dimensional
quantum gravity in the framework of causal dynamical triangulations
can be viewed as arising through a stochastic quantization. 
This requires that the proper time appearing
in the string field theory be identified with the stochastic
time of the stochastic formulation. The framework of stochastic quantization 
gives rise to a natural nonperturbative quantum Hamiltonian, which incorporates
a sum over all spacetime topologies. We point out that the external character
of stochastic time is a feature that pertains more generally to the proper time or distance
appearing in nonperturbative correlation functions in quantum gravity.

\vspace{12pt}
\noindent

%\vfill

\newpage

\section{Introduction}\label{sec1}

{\it Time} is a much-discussed and somewhat enigmatic quantity in classical 
and even more so in quantum general relativity, where
the reparametrization invariance adds to 
the problem of quantizing the theory. Any attempt 
to shed additional light on the role of time in a quantized theory 
of gravity is therefore of interest. Because some of the structural issues
concerning time persist also in two spacetime dimensions, one may profitably
study toy models of two-dimensional quantum gravity to learn about their
resolution. The group of spacetime diffeomorphisms still acts in an analogous
fashion to that in four-dimensional general relativity, while the quantization
can be carried out without having 
to deal with the problem of perturbative nonrenormalizability 
present in the higher-dimensional, physical theory.

The present piece of work is concerned with the two-dimensional quantum 
gravity known as Lorentzian quantum gravity or quantum gravity 
based on {\it causal dynamical triangulations} (CDT). The name refers to 
the regularization in terms of dynamical, triangulated lattices of the curved spacetimes
appearing in the quantum field theory, when formulated as a nonperturbative 
path integral in Lorentzian signature \cite{ajl2d,cdtgeneral}. It turns out that 
in two dimensions a continuum limit can be taken analytically. 
In this paper we will assume this has been done, and work exclusively 
with the resulting continuum theory.
We will show that the string field theory we have developed earlier \cite{cdt-sft}
for the purpose of describing the splitting and joining in time of 
spatial universes has
a natural description as a stochastic quantization of space.
Recall that the original (and strictly causal) CDT quantization 
employs a global proper-time foliation, with respect to which spatial 
topology changes are forbidden. Generalizing this set-up
by allowing isolated causality-violating points,
space {\it can} now split into disconnected components, 
which may or may not join again at a later time, depending on what
processes the model should incorporate.
In a (quantum-) gravitational theory, where geometry is defined intrinsically, 
this raises interesting questions about 
the global nature of this proper-time variable. 
We showed in previous work that consistency relations hold among
the simply connected quantum amplitudes of the 
two-dimensional theory, which indicates that a global time interpretation
may persist in more complicated
situations involving topology change \cite{cap,cdt-sft}. 

Here we will demonstrate that the stochastic
quantization coincides with the string field theory, and therefore 
that the global proper time has a natural re-interpretation as the 
stochastic time arising in a stochastic quantization of (one-dimensional) 
space.\footnote{Note that this differs from a ``standard"
stochastic quantization of gravity, where stochastic time would appear
{\it in addition} to the time already present as part of the
space{\it time} geometry.}
This phenomenon is not unique to the CDT string field theory, but
was first observed in \cite{sto-kawai} in the context of a 
string field theory developed for
noncritical strings \cite{sft}, after which the CDT construction is modelled.
The relation with stochastic quantization was also found independently 
in a reformulation of matrix models as collective field theories \cite{jevicki},
indicating that we are dealing with a more general phenomenon.
Contrary to the rather intricate way in which it enters in two-dimensional 
{\it Euclidean} quantum gravity (equivalently, noncritical string theory),
the relation is much more straightforward in the case 
of the Lorentzian CDT string field theory. As we will see in the following, it 
can be put to use in a constructive manner
to find a number of quantum observables nonperturbatively, in the sense of being able to
evaluate them on a sum 
over all genera of two-dimensional spacetime.

Of course, the CDT formulation is primarily geared towards solving 
four-dimen\-sio\-nal quantum gravity. In this case the model
cannot be solved analytically, but is being investigated
by computer simulations, which have already led to a number of new and 
interesting results \cite{highd}. Among them are strong indications that
the infrared limit of the 
theory is just that of classical general relativity. Details of the 
ultraviolet limit are still under investigation. Candidates for possible UV completions
still within a field-theoretical framework are
(i) the asymptotic safety scenario with a nontrivial UV 
fixed point \cite{uv,weinberg}, (ii) the scale-invariant 
gravity model advocated by Shaposhnikov et al. \cite{shaposh1,shaposh2},
and (iii) the model of Lifshitz gravity suggested
by Ho\v rava \cite{horava}. To the extent they can be compared,
the structural set-up of the latter is reminiscent of that of the CDT approach: 
one also works with an explicit, global time foliation, and the infrared limit is that of 
general relativity, while the UV limit -- assuming it exists -- 
is highly nonclassical and apparently undergoes a ``dynamical dimensional reduction"
(also observed in \cite{lauscher}).  
Interestingly, the construction of the anisotropic Lifshitz gravity models also bears a
structural resemblance with that of stochastic quantization, 
a fact already noted by Ho\v rava \cite{private}.

The rest of the paper is organized as follows:
in Section \ref{sec2} we review briefly the formalism 
of stochastic quantization, closely following reference \cite{zinn}. 
In Section \ref{sec3} we introduce
the CDT string field theory and show that it can be viewed 
as stochastic quantization of space, if CDT proper time 
is identified with stochastic time. In Section \ref{sec4}
we derive the corresponding nonperturbative Hamiltonian and discuss 
its properties and interpretation.
Section \ref{sec5} summarizes our results and their possible
implications for the nature of time in quantum gravity..

\section{Stochastic quantization}\label{sec2}

This section summarizes the key steps of the stochastic quantization formalism; 
for more extended textbook treatments see, for example, \cite{zinn,chai}.
The Langevin stochastic differential equation for a single variable
$x$ reads 
\beq\label{x2.1}
\dot{x}^{(\n)}(t) = - f\Big(x^{(\n)}(t)\Big) + \sOm \;\n(t),
\eeq
where the dot denotes differentiation with respect to stochastic time $t$,
$\n(t)$ is a Gaussian white-noise term of unit width 
and $f(x)$ a drift force.
We will only con\-sider the case of dissipative diffusion where the 
drift force is conservative, that is,
\beq\label{x2.2}
f(x) =  \frac{\prt S(x)}{\prt x}
\eeq
for some function $S(x)$, a property which ensures the stochastic process satisfies
the principle of detailed balance (see e.g. \cite{zinn}).
Without the noise term, \rf{x2.1} reduces to a relaxation equation.
In that case -- depending on the initial value $x_0=x(0)$ -- $x(t)$ will move towards
the ``nearest''  local minimum of $S(x)$ or run away if there is no
minimum which can be reached from $x_0$ by decreasing $S(x)$. When the 
noise term is added, $x(t)$ will be kicked around close to a minimum. If 
there are several local minima, the noise term can kick it from
one to another and also to a region of no minimum if it exists.
In this manner the noise term creates a probability distribution of $x(t)$, reflecting
the assumed stochastic nature of the noise term, with an associated 
probability distribution 
\beq\label{x2.3}
P(x,x_0;t) = \lla \del (x-x^{(\n)}(t;x_0))\rra_\n,
\eeq
where the expectation value refers to an average over the Gaussian noise.
It can be shown that $P(x,x_0;t)$ satisfies the so-called 
Fokker-Planck equation
\beq\label{x2.5}
\frac{\prt P(x,x_0;t)}{\prt t} = 
 \frac{\prt}{\prt x}\left( \oh\Om \frac{\prt P(x,x_0;t) }{\prt x} + 
f(x) P(x,x_0;t)\right).
\eeq
This is an imaginary-time Schr\"{o}dinger equation, with
$\sOm$ playing a role similar to $\hbar$. It enables us to write
$P$ as a propagator for a Hamiltonian operator $\hH$,
\beq\label{x2.6}
P(x,x_0;t) = \la x | \e^{-t\hH}|x_0\ra,~~~\hH= \oh \Om \hp^2 +i \hp \, f(\hx),
\eeq
with initial condition $x(t=0) = x_0$, and $\hp = -i \prt_x$. It follows that
by defining
\beq\label{x2.6a}
\tG(x_0,x;t) \equiv \frac{\prt}{\prt x_0}\, P(x,x_0;t) 
\eeq
the function $\tG(x_0,x;t)$ satisfies the differential equation
\beq\label{x2.6b} 
\frac{\prt \tG(x_0,x;t)}{\prt t} = 
 \frac{\prt}{\prt x_0}\left( \oh \,\Om\, \frac{\prt \tG(x_0,x;t)}{\prt x_0} - 
f(x_0)\;\tG(x_0,x;t)\right).
\eeq
An explicit example, relevant to the further development of the paper, is given by 
\beq\label{x2.20}
S(x) = \frac{x^3}{3} -\lam x.
\eeq
This polynomial function has a local minimum at $x= \sla $, a local maximum at
$x = -\sla$ and is unbounded from below when $x\to -\infty$. It follows that in
absence of the noise term -- corresponding to the classical, 
unquantized system -- the point
$\sla$ is an attractive fixed point for the classical equation \rf{x2.1} since for all 
$x_0 > -\sla$, $x(t)$ will approach $\sla$ as 
$t \to \infty$. For $x_0 < -\sla$ we have a run-away solution and 
$x(t) \to -\infty$ in a finite time. Omitting the noise term corresponds
to taking the limit $\Om \to 0$. One can then drop the functional average
over the noise in \rf{x2.3} to obtain
\beq\label{x2.21}
P_{cl}(x,x_0;t) = \del(x-x(t,x_0)),~~~~\tG_{cl}(x_0,x;t) = 
\frac{\prt}{\prt x_0} \del(x-x(t,x_0)).
\eeq
It is readily verified that these functions satisfy 
eqs.\ \rf{x2.5} and \rf{x2.6b} with $\Om =0$, for instance,
\beq\label{x2.22} 
\frac{\prt \tG_{cl}(x_0,x;t)}{\prt t} = 
 \frac{\prt}{\prt x_0} \Big( (\lam - x_0^2 )\, \tG_{cl}(x_0,x;t)\Big).
\eeq

\section{Quantum dynamics of 2d causal triangulations}\label{sec3}

Quantum gravity defined through causal dynamical triangulations 
aims to construct and evaluate the nonperturbative, Lorentzian 
path integral over spacetime geometries $[g_{\m\n}]$, with or without matter
coupling. In dimension two, and assuming we already have performed
a rotation to Euclidean signature (this is well defined in CDT), this approach gives
a definite meaning to the formal (Euclideanized) sum over histories
\beq\label{2.1}
Z(G_{\rm N},\lam)= \int \cD [g_{\m\n}] \; \e^{-S[g_{\m\n}]},
\eeq
where the (Euclidean) Einstein-Hilbert action is given by
\beq\label{2.2}
S[g_{\m\n}] = -\frac{1}{2\pi G_{\rm N}} \int d^2\xi \sqrt{\det{g_{\m\n}}}\;R 
+ \lam \int d^2\xi \sqrt{\det g_{\m\n}},
\eeq
with Newton's constant $G_{\rm N}$ and the cosmological constant $\lambda$.

One thus proceeds in several steps: 
first the CDT lattice regularization is used to define the path integral,
still with Lorentzian signature.
Next, a rotation to Euclidean
signature is performed at the level of the individual triangulations.
We refer the reader to the original articles \cite{al,ajl2d}
or the recent review \cite{zohren} for details.
The resulting real, Euclidean path integral of the form \rf{2.1}
will however differ from a standard one since we insist as part of the kinematical set-up
that each 
path (spacetime history) possess a global time-foliation.\footnote{This is a ``remnant"
of the corresponding structure in the original Lorentzian spacetimes, which ensures
they are well-behaved causally.}
One then performs a continuum limit by shrinking the individual triangular building blocks to  
zero size (``removing the regulator"), 
while tuning the coupling constant(s) appropriately. This can be done
analytically in the original, strictly causal CDT quantum gravity model. Key quantities 
one can compute in the limit and which contain information about the underlying
quantum geometry of this continuum theory are so-called ``loop amplitudes".
An important example is the amplitude denoted by $G_0(l_0,l;t)/l_0$
that (one-dimensional, compactified) space has length $l_0$ at proper time $t=0$ and 
length $l$ at a later proper time $t$.
The quantity $G_0(l_0,l;t)$ without the normalization factor $1/l_0$ 
has the same interpretation as a transition
amplitude, but with a distinguished marked point on the initial spatial loop 
$l_0$ (the marking removes the symmetry factor $1/l_0$). 
It is convenient to introduce the Laplace transform $\tG_0$ of $G_0$ by
\beq\label{y2.3}
\tG_0 (x_0,x;t) = \int_0^\infty \d l_0  \int_0^\infty \d l
\; \e^{-x_0l_0-xl}G_0 (l_0,l;t),
\eeq
where the variables $x_0$ and $x$ can be interpreted as boundary cosmological 
constants. In the original paper on two-dimensional 
CDT quantum gravity \cite{al} it was shown that 
$\tG_0(x_0,x;t)$ satisfies the 
differential equation
\beq\label{y2.4}
\frac{\prt \tG_0(x_0,x;t)}{\prt t} = 
 \frac{\prt}{\prt x_0}\left((\lam-x_0^2) \; \tG_0(x_0,x;t)\right).
\eeq
Note that (up to a minus sign) $\tG_0(x_0,x;t)$ is obtained from the Laplace transform
of $G_0(l_0,l;t)/l_0$ by differentiating with respect to $x_0$, in the same 
way as $\tG(x_0,x;t)$ in eq.\ \rf{x2.6a} was obtained from $P(x,x_0;t)$.

Comparing now eqs.\ \rf{y2.4} and \rf{x2.22}, we see that we can formally re-interpret 
$\tG_0(x_0,x;t)$ -- an amplitude obtained by nonperturbatively quantizing Lorentzian
pure gravity in two dimensions -- 
as the ``{\it classical} probability'' $\tG_{cl}(x_0,x;t)$ corresponding
to the action $S(x) = -\lam x +x^3/3$ of a zero-dimensional system
in the context of stochastic quantization. 
Stochastic quantization of the
system amounts to replacing
\beq\label{y2.5}
\tG_0 (x_0,x;t) \to \tG(x_0,x;t),
\eeq
where $\tG(x_0,x;t)$ satisfies the differential equation
corresponding to eq.\ \rf{x2.6b}, namely,
\beq\label{y2.6}
\frac{\prt \tG(x_0,x;t)}{\prt t} = 
 \frac{\prt}{\prt x_0}\left( g \frac{\prt}{\prt x_0} + \lam-x_0^2\right) 
\tG(x_0,x;t).
\eeq
For reasons which will become apparent below, we have introduced the
parameter $g:=\Om/2$. Before turning to the physical 
interpretation of eq.\ \rf{y2.6}, let us calculate the so-called 
Hartle-Hawking wave function, which 
in the CDT {\it string field theory}\footnote{i.e. isolated branchings and mergings are now
allowed} is defined as 
\beq\label{y2.7}
\tW(x_0) = \int_0^\infty \d t \;\tG(x_0,l=0;t).
\eeq
The integrand $\tG(x_0,l;t)$ is obtained from $G(l_0,l;t)$ by making a 
Laplace transformation (as in \rf{y2.3}) only of $l_0$ and not of $l$. By construction,
$\tW(x_0)$ accounts for all spacetime histories starting with a single spatial loop of any
length and ending in ``nothing" (the loop of length zero) at an arbitrary later time. 
For physical reasons we demand that the solution to \rf{y2.6}
should obey
\beq\label{y2.8}
G(l_0,l;t=0) = \del(l-l_0),~~~~G(l_0,l;t=\infty)=0,
\eeq
conditions which also hold for the pure-gravity amplitude $G_0(l_0,l;t)$
without topo\-logy change.
Integrating relation \rf{y2.6} from time $t=0$ to infinity we obtain
\beq\label{y2.9}
-1 =   \frac{\prt}{\prt x_0}\left( g \frac{\prt}{\prt x_0} +\lam- x_0^2\right) 
\tW(x_0).
\eeq
This is precisely the differential equation for $\tW(x_0)$ obtained
recently \cite{genera} from a matrix model representation of CDT string
field theory if $g$ was identified with the string coupling
constant, associated with the merging or splitting of spatial
universes as a function of time $t$. Just as in the original pure-gravity CDT model,
the parameter $t$ in the string field theory
was identified with proper time. We now see that within the extended CDT framework,
where topology change is allowed, this time acquires a new interpretation as
{\it stochastic time} and the CDT string field theory that of a stochastic quantization.
 
Note that eqs.\ \rf{y2.6} and \rf{y2.9} are highly nonperturbative in the sense of 
describing a third-quantized
system of geometry, incorporating topology changes of space. 
Eq.\ \rf{y2.9} for $\tW(x_0)$ was originally derived in a matrix model 
representation of the CDT string field theory.\footnote{for a rescaled
version in terms of dimensionless parameters; c.f. eq.\ (30) of ref. \cite{genera}}
What we have done here is to derive these expressions  
by applying ``blindly" the rules of stochastic quantization, treating ``$x$" 
as an ordinary variable, like the position of a particle, whereas
in reality $x$ is the boundary cosmological constant introduced
by the Laplace transformation \rf{y2.3}. A variable with a more direct physical
interpretation is the conjugate length variable $l$ of the boundary, measuring
the size of the spatial universe. The Hamiltonian as a function
of this physical length can be obtained 
by an inverse Laplace transform from the ``classical''
Hamiltonian $\hH_0 = -\d/\d x (\lam-x^2)$ from \rf{x2.6} with $\Om =0$, leading to
\beq\label{3.8}
\hH_0(l) = -l \frac{\d^2}{\d l^2} +\lam \;l.
\eeq
This is a standard Hermitian operator on wave functions $\psi(l)$
on the positive real axis, which are 
square-integrable with respect to the scalar product
\beq\label{3.9}
\lla \psi_1 | \psi_2\rra = \int_0^\infty \frac{\d l}{l}\;\psi^*_1(l) \psi_2(l).
\eeq
The scalar product is fixed uniquely by requiring appropriate 
composition properties 
of the propagator $G_0(l_0,l;t)$ \cite{ajl2d}.
The eigenfunctions $\psi_n(l)$ of $\hH_0(l)$ are the states of 
the spatial universe which are propagated unchanged by $\hG_0= \e^{-t\hH_0}$
with kernel $\hG_0(l_1,l_2;t)$. When we Laplace-transform this 
to $x$-space the scalar product to be used is the one inherited
from $l$-space. For instance, the Laplace transform of 
$\del (l_1-l_2)$ is $1/(x+y)$, which acts like the appropriate 
$\del$-function in $x$-space. In other words, the physically motivated
boundary conditions are different from the ones one would choose if $x$ were
a standard configuration space variable. Likewise, an acceptable eigenfunction 
of $\hH_0(x)$ is not a standard square-integrable function on the real $x$-axis. 
Consequently, instead of \rf{x2.21} we should use
\beq\label{3.10}
\tG_0(x_0,x;t) = \frac{\d }{\d x_0} \left(\frac{1}{x(t,x_0)+x}\right) = 
-\frac{x^2(t,x_0)-\lam}{x_0^2-\lam} \; \frac{1}{(x(t,x_0)+x)^2}.
\eeq

Despite these differences compared
to the situation in ``ordinary'' $x$-space the formal derivation of 
stochastic quantization is unchanged. A neat geometric interpretation
of how stochastic quantization can capture topologically nontrivial
amplitudes has been given in \cite{sto-kawai}. Applied to the present
case, we can view 
the propagation in stochastic time $t$ for a given noise term $\n(t)$ 
as classical in the sense that solving the 
Langevin equation \rf{x2.1} for $x^{(\n)}(t)$ iteratively gives
precisely the tree diagrams with one external leg 
corresponding to the action $S(x)$ (and including the derivative
term $\dot x^{(\n)}(t)$), with 
the noise term acting as a source term. Performing the functional 
integration over the Gaussian noise term corresponds to integrating out the 
sources and creating loops, or, if we have several independent trees,
to merging these trees and creating diagrams with several external legs.
If the dynamics of the quantum states of the spatial universe 
takes place via the strictly causal CDT-propagator $\hG_0 = \e^{-t \hH_0}$, 
a single spatial universe of length $l$ 
cannot split into two spatial universes. Similarly, no two spatial universes are 
allowed to merge as a function of stochastic time.
However, introducing the noise term {\it and} subsequently
performing a functional
integration over it makes these processes possible. 
This explains how the stochastic quantization can automatically generate
the amplitudes which are introduced by hand
in a string field theory, be it of Euclidean character as described
in \cite{sto-kawai}, or within the framework of CDT.

What is new in the CDT string field theory considered
here is that we can use the corresponding 
stochastic field theory to solve the model, since we arrive at 
closed equations valid to all orders in the genus expansion.
Equations \rf{y2.6} and \rf{y2.9} are such examples.
Translating them to $l$-space and using the boundary conditions
$W(l=0)=1$ and $G(l_0=0,x;t)=0$ (because the loop of length $l_0$ is marked),
we obtain from \rf{y2.9} 
\beq\label{3.11a}
\hH(l) W(l)=0,
\eeq
which is a Wheeler-deWitt type equation for the spatial universe.
In addition, we have 
\beq\label{3.11}
 \frac{\prt G(l_0,l;t)}{\prt t} =-\hH(l_0)\, G(l_0,l;t),
\eeq
where the {\it extended} Hamiltonian
\beq\label{3.12}
\hH(l) = -l \frac{\prt^2}{\prt l^2} +\lam l - g l^2
\eeq
now has an extra potential term coming from the inclusion of branching 
points.\footnote{Since in the derivation of $\hH(l)$ we considered only
loop-loop amplitudes (as opposed to arbitrary multi-loop amplitudes), 
this Hamiltonian seems to capture only a  sector of the full dynamics of the
string field theory.
To what extent $\hH(l)$ already incorporates the complete dynamics in some
``effective" way -- as suggested by the fact that
it does contain an infinite genus summation --
is an issue that remains to be understood better.}
Eq.\ \rf{3.11a} is readily solved in terms of the Airy function $Bi$, namely,
\beq\label{3.14}
W(l) =  \frac{{\rm Bi}\Big(\frac{\lam}{g^{2/3}}-g^{1/3} l\Big)}{
{\rm Bi}\Big(\frac{\lam}{g^{2/3}}\Big)},
\eeq
while
\beq\label{3.13}
G(l_0,l;t) = \la l | e^{-t \hH(l)}|l_0\ra
\eeq
describes the nonperturbative propagation of a spatial loop 
of length $l_0$ to a spatial loop of length $l$ in proper 
(or stochastic) time $t$, now including the summation over all genera.
The Hamiltonian $\hH(l)$ is a well-defined Hermitian operator with respect to 
the measure \rf{3.9}. 

\section{The extended Hamiltonian}\label{sec4}

Let us recap the results of the original CDT model, where space was not
allowed to split into disconnected parts \cite{al,ajl2d,zohren}.
We have a Hamiltonian $\hH_0(l)$ and a corresponding eigenvalue equation 
\beq\label{4.1}
\hH_0(l) \psi_n(l)= E_n \psi_n^{(0)}(l),~~~
\hH_0 (l)=  -l \frac{\prt^2}{\prt l^2} +\lam l.
\eeq 
The eigenfunctions and eigenvalues are given by
\beq\label{4.1a}
\psi^{(0)}_n(l) = p_n(l)\; \e^{-\sla l},~~~~E_n= 2n \sla,~~n=1,2,\ldots,
\eeq 
where the $p_n(l)$ are polynomials in $\sla l$ and $p_n(0)=0$. 
Furthermore, we have 
\beq\label{4.2}
\hH_0(l) W_0(l)=0,~~~~W_0(l) = \e^{-\sla l},
\eeq
where $W_0(l)$ is the Hartle-Hawking wave function of the 
original CDT model and relations \rf{4.2} are the counterparts of
\rf{3.11a} and \rf{3.14} when $g=0$. Formally, the amplitude $W_0(l)$ is a 
solution to eq.\ \rf{4.1}
with eigenvalue $E=0$. However, $E=0$ does not belong to the spectrum of 
$\hH_0$ since $W_0(l)$ is not integrable at zero with respect to the measure
\rf{3.9}. Exactly the same is true for the extended Hamiltonian $\hH(l)$ and 
the corresponding Hartle-Hawking amplitude $W(l)$. --
In order to analyze the spectrum of $\hH(l)$,
it is convenient to put the differential operator into standard form.
After a change of variables
\beq\label{4.3}
l= \oh z^2,~~~~~\psi(l) = \sqrt{z} \phi(z),
\eeq
the eigenvalue equation becomes
\beq\label{4.4}
\hH(z)\phi(z) = E \phi(z),~~~~\hH(z) = -\oh \frac{\d^2}{\d z^2} 
+\oh \lam z^2 + \frac{3}{8z^2}-\frac{g}{4} z^4.
\eeq 
This shows that the potential is unbounded from below, but 
such that the eigenvalue spectrum is still discrete\footnote{Whenever
the potential is unbounded below with fall-off faster than $- z^2$, the spectrum is discrete,
reflecting the fact that the classical escape time to infinity is finite.
In this way, the unbounded potential behaves effectively like a finite box.
In addition, like in the case of a box, there exists a one-parameter 
family of selfadjoint Hamiltonians, depending on the specific 
choice of boundary condition one imposes at infinity. See reference \cite{ak} 
for a more detailed discussion relevant
to the present situation.}. For small $g$, 
there is a large barrier of height $\lam^2/(2g)$ 
separating the unbounded region 
for $l > \lam/g$ from the region $0 \leq l \leq \lam/(2g)$ where the 
potential grows. This situation is perfectly suited to applying a standard WKB 
analysis. For energies less than $\lam^2/(2g)$, the eigenfunctions
of $\hH_0(l)$ will be good approximations to those of $\hH(l)$. 
However, when $l > \lam/g$ the exponential fall-off of $\psi_n^{(0)}(l)$
will be replaced by an oscillatory behaviour, with the wave function falling 
off only like $1/l^{1/4}$. The corresponding $\psi_n(l)$ is still
square-integrable since we have to use the measure \rf{3.9}.
For energies larger than  $\lam^2/(2g)$, the solutions will be 
entirely oscillatory, but still square-integrable.

What follows from our analysis is that a
dramatic change has occurred in the quantum behaviour of the one-dimensional
universe as a consequence of allowing topology changes.
In the original, strictly causal quantum gravity model 
an eigenstate $\psi_n^{(0)}(l)$ of the spatial universe had an average size
of order $1/\sqrt{\lam}$, increasing as a function of energy.
Allowing for topology changes (and assuming $g$ suitably small and $n$ not too large),
only the large-$l$ tail of $\psi_n^{(0)}(l)$ will change.
As a result, the probability $|\psi_n(l)|^2/l$ for finding a universe with size
in the interval $[l,l+dl]$ is almost unchanged as long
as $l < \lam/g$. However, the average size of the universe is now infinite!
We see now that the oscillatory behaviour of the amplitude
$W(l)$ for $l > \lam/g$ already observed in \cite{genera}
can be understood as a consequence of $l$ lying in the 
region where the potential in $\hH(l)$ is unbounded below.

We still need to choose a selfadjoint extension
of $\hH(l)$ such that the spectrum of $\hH(l)$ can be determined unambiguously.
One way of doing this is to appeal again to stochastic quantization, 
following the strategy used by Greensite and Halpern \cite{gh}, which was applied to 
the double-scaling limit of matrix models in \cite{ag1,ag2,ak}.
The Hamiltonian \rf{x2.6} corresponding to the Fokker-Planck equation \rf{y2.6}, namely,
\beq\label{4.5}
\hH(x)\psi(x) = -g \frac{\d^2 \psi(x)}{\d x^2} +\frac{\d}{\d x}
\left(\frac{\d S(x)}{\d x}\, \psi(x) \right),~~~~ 
S(x) = \left(\frac{x^3}{3}-\lam\,x\right),
\eeq  
is not Hermitian if we view $x$ as an ordinary real 
variable and wave functions $\psi(x)$ as endowed with the standard scalar
product on the real line. However, by a similarity transformation one can transform $\hH(x)$
to a new operator 
\beq\label{4.6}
\tH(x) = \e^{-S(x)/2g}\hH(x) \, \e^{S(x)/2g};~~~\tpsi(x) = \e^{-S(x)/2g}\psi(x),
\eeq 
which {\it is} Hermitian on $L^2(R,dx)$.
We have 
\beq\label{4.7}
\tH(x)= -g\frac{\d^2}{\d x^2} +
\left(\frac{1}{4g} \left(\frac{\d S(x)}{\d x}\right)^2+
\oh \frac{\d^2 S(x)}{\d x^2}\right),
\eeq
which after substitution of the explicit form of the action becomes
\beq\label{4.7a}
\tH(x)=-g \frac{\d^2}{\d x^2} +V(x),~~~~V(x)= \frac{1}{4g} (\lam -x^2)^2+ x.
\eeq
The fact that one can write 
\beq\label{4.8}
\tH(x)=  \hR^{\dg}\hR,~~~~
\hR=-\sqrt{g}\frac{\d}{\d x} +\frac{1}{2\sqrt{g}}\frac{\d S(x)}{\d x}
\eeq
implies that the spectrum of $\tH(x)$ is positive, discrete and 
unambiguous. We conclude that the formalism of stochastic quantization
has provided us with a nonperturbative definition of the CDT 
string field theory.

\section{Summary and discussion}\label{sec5}

In this paper we have shown that there is an alternative derivation, using
stochastic quantization, of the CDT string
field theory introduced earlier in \cite{cdt-sft,genera}.
The stochastic quantization is not performed for the initial
path integral over all spacetime geometries, but at the level of
the effective continuum dynamics of the spatial geometry of the
universe, which for the case of gravity in 1+1 dimensions is described by
a single variable, the universe's size or length. Interestingly, in order
for the equivalence to hold, we had to identify the stochastic time
of the construction with the proper time of the original CDT model. 
As a bonus, the stochastic quantization naturally led us to {\it a}
nonperturbative definition of the CDT string field theory.
This is nontrivial, because the theory contains a
sum over all spacetime topologies.
Our construction mirrored that of Kawai et al. \cite{sto-kawai}, who were
the first to observe (in a Euclidean context) that
the noncritical string field theory developed by them
could be viewed as a stochastic quantization of space, with
stochastic time playing the role of proper time in the 
corresponding two-dimensional quantum gravity theory.
Physically, the two string field theories are of course different. In the CDT case
we were able to push the formalism further to obtain an explicit quantum
Hamiltonian and analyze its spectral properties.

At first sight, it may seem curious that stochastic time -- usually thought
of as a fictitious, external parameter -- makes an appearance as the ``time"
of a quantum-gravitational theory. However, it may be argued that 
the external character of this particular {\it distance parameter}
is something found more generally in the construction of diffeomorphism-invariant
correlation functions in nonperturbative quantum gravity.
As a simple example, consider the case of a scalar field $\phi$ coupled to (Euclidean)
quantum gravity in two dimensions.
A diffeomorphism-invariant definition of a two-point
correlator can be obtained by integrating over all pairs of insertion points
of the matter fields which are a geodesic distance $R$ apart, that is,
\bea\label{5.1}
\lefteqn{\la \phi(R)\phi(0)\ra = 
\int \cD [g_{\m\n}(\xi)]\int \cD\phi(\xi) \; \e^{-S(g_{\m\n},\phi)} \;}  \\ 
&&\times \int \d^2\xi_1 \sqrt{\det g(\xi_1)}\int\d^2\xi_2 \sqrt{\det g(\xi_2)}\; \;
\phi(\xi_1)\phi(\xi_2) \;\del(D_g(\xi_1,\xi_2)-R).
\eea  
The function $D_g(\xi_1,\xi_2)$ appearing in the argument of the $\delta$-function
denotes the geodesic distance between the points labelled $\xi_1$ and $\xi_2$.
As indicated by the notation, this distance depends on the other dynamical
field variable, the metric $g_{\m\n}(\xi)$. 

In this construction,
the geodesic distance $R$ is fixed outside the functional integral, and therefore
may be regarded as external. It does not refer to any particular metric, but 
is the geodesic distance in {\it all}
geometries entering in the functional integral simultaneously. From this point of
view it is of course intimately related to the dynamical quantum properties of
the ensemble. In particular, $R$ can have genuine quantum properties, for example,
it can scale noncanonically. The proper time appearing
in the description of the ``world sheets" of the
string field theories has a similar status. It is a notion of time which is
defined invariantly (in this case as the ``geodesic distance to a one-dimensional
boundary"), and superimposed on an ensemble of geometries. 

It is precisely this notion of proper time which in both Euclidean and Lorentzian 
two-dimensional quantum gravity with topology change (a.k.a. string field theory in
zero-dimensional target space) apparently is equivalent to stochastic time. Although in
our present derivation the third-quantized nature of the construction appeared in
an essential way, the argument about the ``external" nature of this time in correlation
functions we made above
appealed neither to the inclusion of nontrivial topology nor the dimensionality of
spacetime. This suggests that stochastic time may play a role in these more
general situations too, a line of enquiry that is currently under investigation.

%Expressions like the one in eq.\ \rf{5.1} highlights the 
%fact that we are in principle dealing with a diffeomorphism 
%invariant ``proper time'' and the stochastic quantization of the CDT model 
%allows us to write down the nonperturbative Hamiltonian corresponding 
%to this time. The eigenstates of this Hamiltonian are eigenstates of the
%spatial universe. The starting point in CDT was a world with Lorentzian 
%signature. In order to perform our calculations we rotated each geometry
%to Euclidean signature in the way described in the original CDT article
%\cite{al}. In that article it was stated that a rotation back to Lorentzian
%signature should be performed at the Hamiltonian level: the proper time
%evolution operator $\e^{-t \hH_0(l)}$ should be replaced by
%$\e^{it \hH_0(l)}$. We can now replace this statement with one
%where the perturbative Hamiltonian $\hH_0(l)$ is replaced by the 
%nonperturbative Hamiltonian $\hH(l)$ and the proper time, which is 
%also the stochastic time, is rotated $t \to -i t_{Lorentz}$, and 
%we have completed our nonperturbative definition of Lorentzian
%quantum gravity. Further implications of this rotation will be 
%discussed elsewhere.

\section*{Acknowledgment}
JA, RL, WW and SZ acknowledge support by
ENRAGE (European Network on
Random Geometry), a Marie Curie Research Training Network in the
European Community's Sixth Framework Programme, network contract
MRTN-CT-2004-005616. RL acknowledges
support by the Netherlands
Organisation for Scientific Research (NWO) under their VICI
program.

%\newpage

\end{document}